\documentclass[notoc]{JHEP}
\usepackage{epsfig}
\usepackage{amsmath}
\usepackage{amssymb}

\def\ra{\rightarrow}

\def\L{{\cal L}}

\def\N{{\cal N}}
\def\O{{\cal O}}
\def\P{{\cal P}}

\def\Z{{\cal Z}}
\def\qslash{\not{\hbox{\kern-2pt $q$}}}
\def\delslash{\not{\hbox{\kern-2pt $\partial$}}}

\def\beq{\begin{equation}}
\def\eeq{\end{equation}}
\def\eeq{\end{equation}}
\def\bea{\begin{eqnarray}}
\def\eea{\end{eqnarray}}
\def\bq{\begin{quote}}
\def\eq{\end{quote}}
\def\lesssim{\mathrel{\mathpalette\vereq<}}
\def\gtrsim{\mathrel{\mathpalette\vereq>}}
\def\lsim{\mathrel{\lesssim}}

\def\Ht{{\tilde H}}
\def\htilde{{\tilde h}}

\makeatletter
\def\vereq#1#2{\lower3pt\vbox{\baselineskip1.5pt \lineskip1.5pt
\ialign{$\m@th#1\hfill##\hfil$\crcr#2\crcr\sim\crcr}}}
\makeatother

\title{\center{New Tools for Fermion Masses \\ from Extra Dimensions}}
\author{David~Elazzar~Kaplan \thanks{dkaplan@slac.stanford.edu}
\\ SLAC, Stanford University, Stanford, CA 94309}
\author{Tim~M.~P.~Tait \thanks{tait@anl.gov} \\ High Energy Physics Division,
Argonne National Laboratory, Argonne, IL 60439}

\abstract{We present models in which the observed fermion
masses and mixings are generated by dynamically localizing the three
generations of matter in a flat compact extra dimension.  
We first construct models assuming the hierarchy problem is
addressed by the existence of large extra dimensions, {\em i.e.}
the fundamental scale is not far above a TeV and supersymmetry
is not imposed.  These models are compactified, chiral, and 
don't require fine-tuning to generate the top mass.
Limits on the compactification scale based on flavor-changing neutral 
currents are relaxed compared to those on existing models.
We then supersymmetrize some of these models. Using
$\N=1$ superspace language in extra dimensions, we find space-dependent 
flat directions which can be used to localize fields.  
Finally, we discuss methods of breaking supersymmetry and the
impact of these models on the superpartner spectrum.
}

\preprint{hep-ph/0110126 \\ ANL-HEP-PR-01-081 \\ EFI-01-44 \\ SLAC-PUB-9021}

\begin{document}

\section{Introduction}
\label{intro}
\indent \indent
Measured fermion masses represent a window into ultraviolet physics.
The standard model (SM) can reproduce the observed hierarchy of
masses and mixing angles with a set of dimensionless parameters
(Yukawa couplings) ranging over five orders of magnitude, but
does not explain why such a diverse and interesting pattern exists.

One avenue of exploration of high energy physics would be to find
models in which such hierarchical patterns can be reproduced by a
theory with only ``natural'' couplings, {\it i.e.,} dimensionless parameters
of order unity.  The first success of this type is the Froggatt-Nielsen
mechanism \cite{Froggatt:1979nt} which imposes an additional symmetry
on the SM thereby forbidding most Yukawa couplings.  Yukawa couplings
are generated by higher dimension operators and the spontaneous breaking
of the additional symmetry and are suppressed by powers of the breaking
scale over some fundamental scale.

An interesting orthogonal approach to generating a large hierarchy in
the Yukawas is to use locality rather than symmetries to produce small
dimensionless numbers.  The Arkani-Hamed-Schmaltz (AS) mechanism requires
SM fermion zero modes to be localized at different positions in one
(or more) extra dimension(s) \cite{Arkani-Hamed:2000dc}.  This can be
done by coupling five-dimensional fermions to a scalar field with
a space-dependent vacuum expectation value (VEV).  For Gaussian wave
functions, couplings between fields are
exponentially suppressed for separations of order a few (in units of
the wave function widths).  It has been shown \cite{Mirabelli:2000ks}
that all fermion masses and mixing angles can be reproduced by localizing
all SU(2) doublets and SU(2) singlets at different positions in one extra
dimension, with the Higgs zero mode constant along the extra
dimension.  However, it has been noted that in the five-dimensional
case one cannot accommodate the observed CP violation in the Kaon
system \cite{Branco:2001rb}.  In addition the large top mass requires some
fine-tuning of parameters.  Many interesting variations on this
theme have since appeared in the literature
\cite{Dvali:2000ha,Kaplan:2000av,Dienes:1998vh,Gherghetta:2001kr}.

A complete version of a model of this type should have a four-dimensional
chiral low energy effective theory.  Five-dimensional theories are in
general non-chiral but can be made chiral by choosing the right boundary
conditions \cite{Cheng:2000bg}.
In the next section, we present a simple set of orbifold boundary
conditions which can reproduce the AS model in a compact extra dimension.
The boundary conditions are realized by compactifying on a ${\cal Z}_2$
orbifold and by giving each fermion a different 5d mass which is odd under
the ${\cal Z}_2$, we localize each fermion at a distinct location in the
extra dimension.

We then present simpler models in which the fermions (and in one of
the models, the Higgs boson VEV) are each localized on one of
two orbifold fixed points.
Different Yukawa couplings are generated due to the fact that the
fermion wave functions have different widths.  Their widths are controlled
by their order one couplings to a scalar field, and their location
({\it i.e.,} which orbifold fixed point they are centered about) is governed
by the sign of the coupling\footnote{A supersymmetric model of this sort,
in the case of an anti-de Sitter background, was discussed in 
\cite{Gherghetta:2001kr}.}.
In this scenario (unlike the AS one), the top mass is natural and is a
result of a quark doublet
and a quark singlet having the opposite sign couplings as the other quarks,
localizing them at the Higgs brane.
In addition, $\epsilon_K$ is predicted to be of the right order because
the Yukawa matrices are ``full'' in the sense that there are no negligible
entries.  Finally, the flavor problems common to
these models are ameliorated and thus a lower compactification scale
is allowed.

In Section 3, we promote our models to a supersymmetric framework.  We
use the notation of Arkani-Hamed, Gregoire and Wacker
\cite{Arkani-Hamed:2001tb} to describe supersymmetry in five dimensions.
In the simplest model, zero modes are localized by mass terms which are
odd under the orbifold.  These terms are allowed by all remaining symmetries
in the theory and can be viewed as VEV's of maximally broken gauge symmetries.
We then find flat directions in which a scalar field has a
space-dependent VEV along the extra dimension.
This allows for additional models where chiral
superfields are localized at arbitrary points.  However, it is difficult
to produce viable models of this type because of the extra constraints
of 5d supersymmetry.  These models can be made to work by supplementing them
with a partial Froggatt-Nielsen mechanism.  We also present another
possibility where we compactify on $S^1$, 
and introduce chirality by hand by
inserting a three brane and ``projecting'' the chiral states into the
bulk using the supersymmetric profiles found earlier in the section.

In Section 4 we discuss some of the issues which naturally come up in this
context.  For instance, are there any distinguishing signals from such
models and how should supersymmetry be broken.  For a
high flavor scale, how supersymmetry breaks plays an important roll
in determining whether or not one can find physical evidence of these
theories.
Mediating supersymmetry breaking can be done in an extra dimensional
context, as in gaugino mediated supersymmetry breaking, or can be completely
orthogonal to this flavor mechanism, such as low-scale gauge mediation.

\section{Non-supersymmetric Models}
\label{nonsusy}
\indent \indent
Our first models use an extra dimension to explain the
small Yukawa interactions apparent from the quark and lepton masses
in terms of fermions localized in the extra dimension.
Localizing quarks and leptons may also be helpful to prevent
unacceptably fast proton decay \cite{Arkani-Hamed:2000dc,Adams:2001za}.
We assume a flat background metric,
$\eta^{M N} = {\rm Diag}[+1, -1, -1, -1, -1]$, where the large
Latin characters $M, N,...$ refer to the full 5d vector indices
and space coordinates $x^M = \{ x^\mu, y \}$ are decomposed into
the 4d (uncompactified) subset $x^\mu$ and the compactified direction,
$y$.
Without supersymmetry, our models suffer from the
usual hierarchy and triviality problems of the SM, and thus we would
like the fundamental scale to be of order TeV so that the large
extra dimension solution to the hierarchy problem
\cite{Arkani-Hamed:1998rs,Antoniadis:1998ig} is
applicable\footnote{For example, we might have more than one extra dimension,
with the SM fields seeing only one of the extra dimensions.}.
As we wish to construct flavor by localizing the fermions of the
SM at various points in the extra dimension, it is necessary
that the SM gauge fields live in the full 5d theory.
We begin by constructing models which reproduce the fermion mass spectrum.
We then examine the effects of this new physics on low energy processes
allowing us to put a bound on the size of the extra dimension and of the
fundamental scale.

We work with a compact extra
dimension subject to orbifold boundary conditions, $S^1 / \Z_2$, with the
orbifold fixed points at $y=0$ and $y=\pm L/2$.
The orbifold is essential in order to recover a chiral
theory from the vector-like 5d theory by removing the mirror
partners of the fermion zero modes.  It is further
useful because it can force the VEV of an
odd scalar field to assume a non-trivial profile with respect to the
extra dimension.  The extra dimension is compact, with $-L/2 < y \leq L/2$
and the points $-L/2$ and $L/2$ identified,
but the orbifold constrains the fields in the region $y<0$ to
shadow the fields in the $y>0$ region, and thus the physical dynamics
may be understood to take place in the region $0 \leq y \leq L/2$.
The 5d theory contains a real scalar ``localizer'' field $\phi$ and
a number of fermions $\psi$ (which correspond to the usual SM quarks and
leptons plus their mirror partner degrees of freedom) satisfying orbifold
boundary conditions \cite{Cheng:2000bg},
\bea
\nonumber
\phi( x^\mu, -y) = - \phi (x^\mu, y) &,&
\phi( x^\mu, L/2+y) = - \phi (x^\mu, -L/2+y) , \\
\psi( x^\mu, -y) = \gamma_5 \psi ( x^\mu, y) &,&
\psi( x^\mu, L/2+y) = \gamma_5 \psi ( x^\mu, -L/2+y) .
\label{eq:orbifold}
\eea
The 5d Lagrangian density is,
\bea
\label{eq:5dL}
{\cal L}_5 &=& \overline{\psi} \left[ i \gamma^M \partial_{M}
- \frac{f_\psi}{\sqrt{M_*}} \phi \right] \psi
+ \frac{1}{2} \partial_M \phi \partial^M \phi
- \frac{\lambda}{4 M_*} ( \phi^2 - u^2 )^2 ,
\eea
where we have ignored the gauge interactions as they are unimportant
with respect to localization.  A mass for the fermions is forbidden
by the orbifold transformations (\ref{eq:orbifold}).  The fundamental scale
$M_*$ has been included in the interactions so that the coupling
constants $f_{\psi}$ and $\lambda$ are dimensionless.

In order to estimate reasonable ranges of the parameters such
as $f_{\psi}$ and $\lambda$, it is necessary to make some assumptions
about the underlying theory.
If the underlying
theory at high energies is such that all couplings are strong at the
cut-off, naive dimensional analysis (NDA) would suggest
$f_{\psi} \sim \sqrt{24 \pi^3}$ and $\lambda \sim 24 \pi^3$
at $M_*$ \cite{Chacko:2000hg}.  These estimates also provide an estimate of
where perturbation theory is expected to break down.
We will consider couplings somewhat smaller than those
suggested by NDA, $f_{\psi} \sim \lambda \sim 1$.  Such couplings
could be considered natural for a perturbative underlying theory
containing only one dimensional parameter $M_*$ (see for example
\cite{Arkani-Hamed:2001ca,Hill:2000mu}).  In constructing models,
we invoke sources for bulk fields living on branes (and in
some cases field theories confined to the branes themselves).  We
will assume the underlying theory is such that the branes can be
treated as thin, rigid objects, and further that the
sources living on them (which we presume have some unspecified
dynamical origin) are generated at a high scale, and will not
succumb to back-reaction effects from the bulk fields.

As discussed in \cite{Georgi:2001wb}, the
orbifold boundary conditions clash with the bulk dynamics for $\phi$,
resulting in a non-trivial VEV which can be approximated for
$M_* L \gg \lambda$ by,
\bea
\label{eq:phivev}
\langle \phi \rangle (y) & = & u \:
{\rm tanh} \left[ \beta (-L/2 -y) \right]
{\rm tanh} \left[ \beta y \right]
{\rm tanh} \left[ \beta (L/2 - y)  \right] ,
\eea
where $\beta^2 = \lambda u^2 /2$.
This nontrivial profile for $\langle \phi \rangle$,
inserted into the 5d Lagrangian \ref{eq:5dL}, appears as a mass
for fermion $\psi$ that varies across
the extra dimension, $M_{\psi}(y) = f_{\psi} \langle \phi \rangle (y)$.
Turning to the 4d effective theory, we expand $\psi$ in a 
Kaluza-Klein (KK) tower
\cite{Kaplan:1992bt}
and find expressions for the zero mass wave functions,
\bea
\psi^0_{\pm} (y) = \N_{\pm} {\rm Exp}
\left[ f_\psi \int^y_0 dy^\prime \langle \phi \rangle(y^\prime) \right] ,
\eea
where $\psi^0_+$ is the left-chiral zero mode and
$\psi^0_-$ the right-chiral one.
For $f_\psi u > 0$, this results in wave function
$\psi^0_+$ localized about the orbifold fixed point $y = 0$,
with profile that looks like $1/ {\rm cosh}^{(\alpha/\beta)} [ \beta y]$,
where $\alpha = | f_\psi u |$.
For $f_\psi u < 0$, $\psi^0_+$ has similar profile,
but centered about the other orbifold point, $y= L/2$.  In each case,
the mirror zero mode $\psi^0_-$ is inconsistent with the orbifold conditions,
Eq.(\ref{eq:orbifold}), and thus removed from the spectrum
\cite{Georgi:2001wb}.

In order to simplify our
analysis, we will further consider the case in which
$\lambda u^2 L^2 \gg 1$, for which the domain walls can be approximated
as step functions,
\bea
\label{eq:simplephi}
\langle \phi \rangle (y) & = & u \: \epsilon ( y )
\eea
with,
\begin{eqnarray}
\epsilon (y) = \left\{ \begin{array}{rcr} +1 & ~~~~~ &\frac{L}{2}>y > 0 \\
-1 & ~~~~~ & \frac{-L}{2}<y < 0 \end{array}
 \right.
\eea
with $L/2$ and $-L/2$ identified.
Again we have a single zero mode with profile,
\bea
\label{eq:zmodeprofile}
\psi^0 (y) & = & \sqrt{ \frac{ 2 \alpha}{1 - e^{- \alpha L}}}
\: e^{- \alpha y} ,
\eea
centered at $y=0$ for $f_\psi u > 0$, or the same wave function with
by $y \ra L/2-y$ for $f_\psi u < 0$.
In regards to flavor in the light quark sector
our exponential profile does not actually differ much from
$1/{\rm cosh}$
because the small entries in the Yukawa interaction matrices
are generated by the small
overlap in the tails of the exponentials which do not differ much
from the tails of the $1/{\rm cosh}$ function.  The major difference
is with respect to the large mass matrix entries, namely the top
Yukawa coupling, which come from large overlaps, and thus are
more sensitive to the shape of the entire wave functions.
In that case, as we shall see below, the more narrow
exponential will have greater difficulty realizing an $\O(1)$ top
Yukawa coupling than the broader $1/{\rm cosh}$ would have, and thus
it is somewhat more difficult to realize flavor for our limiting case
than it would be for the general $1/{\rm cosh}$ zero modes.

We now discuss the AS model,
the original proposal to generate flavor in a large extra dimension
\cite{Arkani-Hamed:2000dc}, and construct two explicit
non-supersymmetric models of flavor,
finishing with some remarks on the experimental constraints
on this class of models, and whether they allow the extra dimension to
really solve the hierarchy problem.

\subsection{The Arkani-Hamed-Schmaltz Model}
\label{sec:as}

The AS model generates flavor by localizing zero modes
of the weak doublet and singlet fermions at different positions, with
the Higgs VEV spread evenly throughout the bulk.
The 4d Yukawa interactions arise as the overlap of a doublet with a
singlet field.  It is assumed that the wave functions are Gaussians
with common widths $\alpha$ (presumably of the order of the
fundamental scale $M_*$); flavor is successfully realized
by distributing the fermions appropriately throughout a region of about
$\Delta y \sim 25 / \alpha$, determined from a numerical scan of
parameters by Mirabelli and Schmaltz \cite{Mirabelli:2000ks}.

In order to remove the troublesome mirror fermions, it is desirable to
impose the orbifold on the AS model.  Since we need to have the
fermion zero modes spread (roughly) evenly through the extra dimension,
and to have Gaussian wave functions, we would like the localizer VEV to
be approximately linear.  This can be engineered by including
sources for $\partial_y \phi$ at the orbifold fixed points,
\bea
\label{eq:dsources}
  J_1 \left( \partial_y \phi \right) \delta \left( y \right)
+ J_2 \left( \partial_y \phi \right) \delta \left( y - L/2 \right) .
\eea
If $\phi$ were massless and had no quartic interaction, these sources
would literally result in a linear VEV.  For a massive localizer,
the VEV may be simply obtained by using the Green's function for
a simple harmonic oscillator of imaginary frequency and we find
that when $m \lsim M_* / 3$ (as could be expected if $m^2$ is generated
at one-loop), the VEV remains approximately linear, as demonstrated
in Figure~\ref{fig:localizer}.
In order to have each fermion localized about a different point
in the extra dimension, we introduce ``odd masses''
in $\L_5$ for each fermion,
\bea
\label{eq:orbimass}
\L^M_5 &=& M_{\psi} \;  \epsilon (y) \; \overline{\psi} \, \psi .
\eea
This term could come from the VEV of an second scalar field which is odd
under the $\cal{Z}_2$ orbifold symmetry and has the appropriate sources
at the orbifold fixed points.  The odd mass effectively shifts the
zero crossing of the ``mass function'' for the fermion, thus localizing it
some distance away from one of the orbifold fixed points.

\FIGURE[t]{
\epsfysize=4.0in
\centerline{\epsfbox{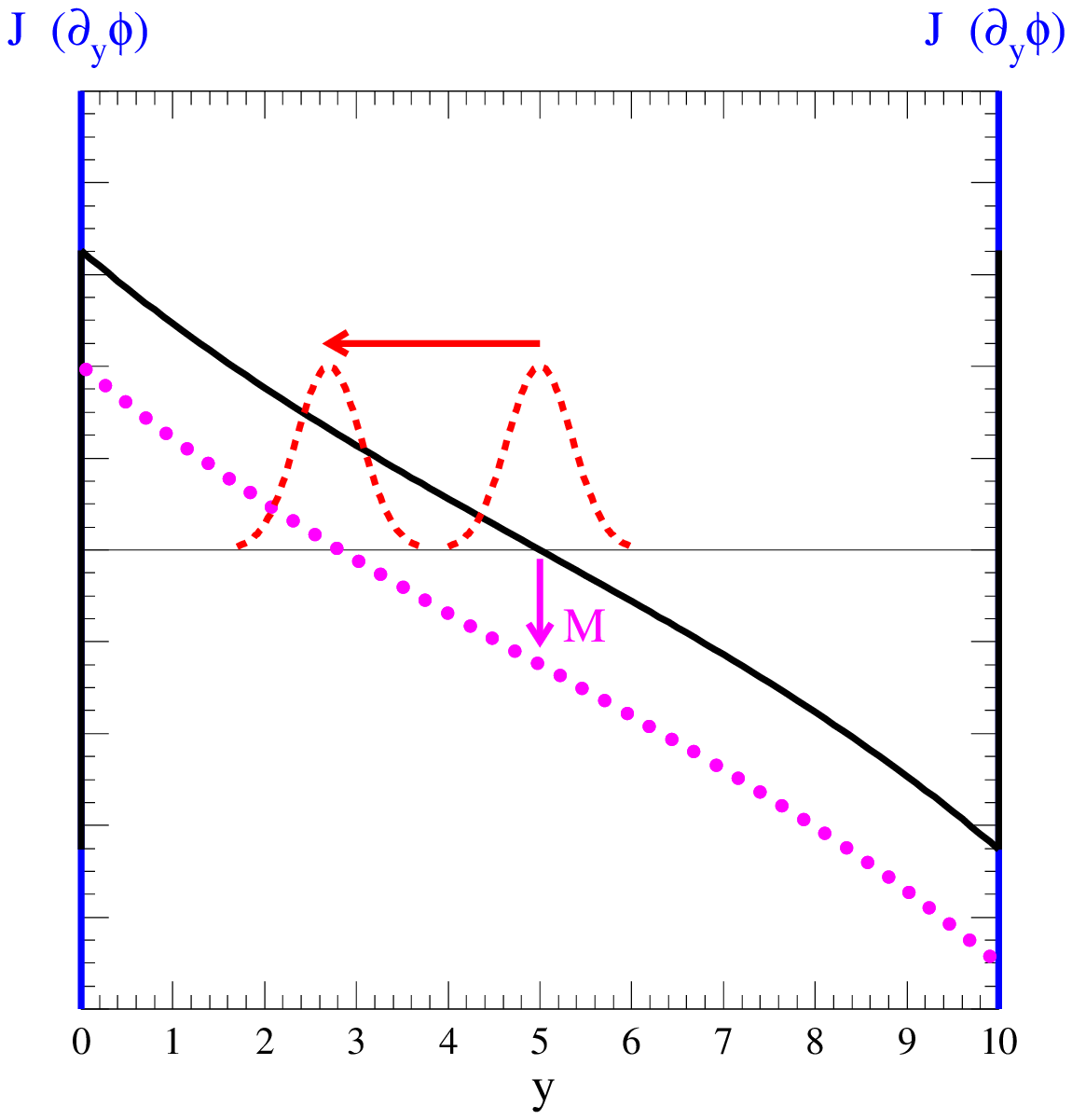}}
\caption{The profile (solid curve) for $\langle \phi \rangle$
resulting from the sources Eq.~\protect{\ref{eq:dsources}}
with $J_1 = J_2$ and the mass of the localizer taken to be
$m \sim M_* / 5$.  Also shown is the effective mass function (dotted curve)
seen by a fermion with odd mass $M \sim J$, and shift in the
wave function (dashed curves)
which results from this odd mass.}
\label{fig:localizer}}

\subsection{Higgs in the Bulk}
\label{sec:bulkhiggs}

In our first model, the Higgs lives in the entire 5d bulk, and is
even under the orbifold transformation.  A 5d version of the
SM Higgs potential
will thus generate an EWSB VEV $v$ spread evenly throughout the extra
dimension.  The underlying 5d Yukawa interactions are,
\bea
\L &=&
  \frac{Y^d_{ij}}{\sqrt{M_*}} H \overline{q}_i d_j
+ \frac{Y^u_{ij}}{\sqrt{M_*}} H^c \overline{q}_i u_j + h.c.
,
\eea
where $H$ is the Higgs doublet, $H^c = i \sigma_2 H^*$,
$q_i$ with $i=1,2,3$ are the three families of weak doublet quarks,
and $u_i$ and $d_i$ are the up- and down-type weak singlet quarks,
respectively.  We have included the appropriate power of the
cut-off  scale for the effective 5d theory, $M_*$,
such that the $Y^u$ and $Y^d$ are dimensionless.
Again, we assume that all of the $Y^u$ and $Y^d$
are of
$\O(1)$ (though not necessarily identically equal to one, and with $\O(1)$
complex phases with respect to one another).

We realize a hierarchy in the
effective 4d theory by coupling the weak doublets to $\phi$ with
couplings $f_{q_i} > 0$ and the weak singlets
to $\phi$ with $f_{u_i}, f_{d_i} < 0$.
This results in the doublet zero modes centered at
$y=0$ with exponential widths $\alpha_i$ while the singlet
zero modes (both up- and down-type) are centered at $y=L/2$,
again with a variety of widths.
Provided $u \sim M^{3/2}_*$, $\lambda \sim 1$, and $f_i \sim 1$,
the widths $\alpha$ will be of order $M_*$.
The effective coupling strength between the Higgs and the
zero modes of the left-handed doublet $i$ and right-handed singlet $j$ are,
\bea
\label{eq:4dyukawas}
y_{ij} &=& \N_{ij} \: \frac{Y_{ij}}{\sqrt{M_* L}} \:
\frac{        e^{\frac{-\alpha_j L}{2}}
            - e^{\frac{-\alpha_i L}{2}} }
{\left( \alpha_i - \alpha_j \right)} ,
\eea
where the normalization factor is given by
$\N^2_{ij} = 4 \alpha_i \alpha_j/[(1-e^{-\alpha_i L})(1-e^{-\alpha_j L})]$.
This equation is valid for both the up- and down-type Yukawa interactions,
with the appropriate $\alpha_j$ for the right-handed field in each case.
The basic idea is that the third family fermions are more weakly coupled
to $\phi$, resulting in a large
overlap between the doublets and singlets, and thus strong coupling to the
Higgs, whereas the second and first generations couple more strongly to
$\phi$, and thus have narrower profiles with exponentially suppressed
overlaps and hence smaller interactions with the Higgs.

The model contains nine parameters (three $\alpha_{q_i}$, three
$\alpha_{u_i}$, and three $\alpha_{d_i}$) and to be considered successful,
must fit the six quark masses and three real CKM angles with all of the
widths of $\O(1)$.
Generally, there is some tension in successfully
generating the flavor observed in nature.  The large top mass
requires that the $u_3$ and $q_3$ zero modes be rather broad, which tends to
generate large entries in the 13 and 23 entries of the mass
matrices.
Working only at the level of order of magnitudes, we find that for
$M_* L = 10$, one can successfully
realize quark flavor with widths ranging from $1/2$ (for $u_3$) to
3 (for $d_2$ and $q_1$), with the $Y_{ij}$ ranging from about 3 (for
$Y^u_{33}$) to $1/3$ (for $Y^{u(d)}_{23}$ and $Y^{u(d)}_{13}$ ).  For
$M_* L = 20$ one has widths from $1/5$ to $3/2$, with the same range of
$Y$.

The lepton sector can be constructed by introducing the lepton doublets,
$l_i$, right-handed charged singlets, $e_i$, and three right-handed
neutrinos $\nu_i$ into the bulk, each coupled to the localizer.  In the
absence of any symmetries to protect them, we assume Majorana masses for
the $\nu$ fields on the order of $M_*$ which we now take to be 100 TeV.
The 5d Yukawa interactions and bulk masses are,
\bea
\L &=&
  \frac{Y_{ij}^e}{\sqrt{M_*}} \; H \; \overline{l}_i \, e_j
+ \frac{Y_{ij}^\nu}{\sqrt{M_*}} \; H^c \;
  \overline{l}_i \, \nu_j + M^\nu_{R ij} \; \nu^c_i \, \nu_j + h.c. \: ,
\eea
where $M^\nu_{R ij} \sim 100$ TeV are the Majorana masses for the
right-handed neutrinos, and need not be diagonal in the same basis
as the interactions with $\phi$.  The interactions of the zero modes
include Yukawa interactions ($y^e$ and $y^\nu$) suppressed by the
overlap of the zero-mode wave functions, as in Eq.(\ref{eq:4dyukawas}).
When the Higgs acquires a VEV, this results in Dirac masses for the
both the charged and neutral leptons.  The charged lepton mass matrix
can be diagonalized as was done for the quarks, but the neutrinos are
more conveniently analyzed by first integrating out the heavy right-handed
neutrinos.  This results in effective Majorana masses for the left-handed
neutrinos,
\bea
\label{eq:mnuL}
M^\nu_{L ij} &=& v^2 y^\nu_{ik} \left( M^\nu_{R} \right)_{kl}^{-1}
y^{\nu *}_{lj} .
\eea

We attempt to understand lepton flavor by building a hierarchy into the
4d Yukawa interactions, $y^e$ and $y^\nu$, arising from the exponentially
suppressed overlaps
of the left- and right-handed lepton wave functions.  We find it is
generically easy to produce a heaviest neutrino relevant for atmospheric
neutrino oscillation by simply arranging the wave functions such that
$y^\nu_{23} \sim y^\nu_{33} \sim 10^{-5}$, which results in a neutrino
with mass $m^2 \sim 10^{-3} {\rm eV}^2$ which is almost an equal mixture
of $\nu_\mu$ and $\nu_\tau$ (some fine-tuning is required for the mixing
to be near maximal).  A small mixing angle solution for the
solar neutrinos may then be produced by introducing much smaller entries
$y^\nu_{11} \sim y^\nu_{12} \sim y^\nu_{22} \sim 10^{-6}$, producing a
neutrino with mass $m^2 \sim 10^{-5} {\rm eV}^2$ which is almost entirely
an electron neutrino, with small $\nu$ and $\tau$ components such that
$\sin^2 \theta \sim 10^{-2}$.  The third neutrino is generally light and
is largely the mixture of $\nu_\tau$ and $\nu_\mu$ orthogonal to the
heaviest neutrino.  This scheme can be realized within the context of
Eq.(\ref{eq:4dyukawas}) when $M_* L = 10$ for widths varying between about
$1$ (for $e_3$) to about $4$ (for $\nu_2$ and $l_1$), and the underlying
$Y$ range between about $3$ and $1/4$.  When $M_* L = 20$, we find that
we need widths between about $1/2$ (for $e_3$) to $5/2$ (for
$\nu_2$ and $l_1$) with the same range of $Y$.  The large mixing angle
solution to the solar neutrino problem is difficult to realize in this
scenario.

\subsection{Localized Higgs VEV}
\label{sec:localhiggs}

Now we present what we believe is the most attractive scenario
for extra-dimensional flavor:  the possibility that the Higgs VEV
is confined to one of the orbifold fixed points.
This could be accomplished in a number of different ways.
One option is that the Higgs field is simply confined to the boundary,
and thus EWSB occurs only at a single point in the extra dimension.
Another idea is that the Higgs is a bulk field, with a positive
bulk ${\rm mass}^2$ such that it does not develop a bulk VEV,
but a separate negative ${\rm mass}^2$ term exists on the boundary,
and again the EWSB VEV develops only close to the boundary.  A final
scenario has the Higgs field in the bulk, coupled to more bulk
fields (such as the localizer $\phi$) which have VEV's which are
functions of $y$ and trigger EWSB only in a limited region of
$y$.  Generally one would expect some fine tuning associated with any
of these options, since both bulk and boundary masses would naturally be of
order $M_* \gg v$.

The fermions will be localized as before
with $f_{\psi} \sim \O (1)$, which again will result in $\O(1)$
widths for their exponential zero mode profiles.
If the Higgs and its VEV are, for example,
confined to $y=0$, the 5d mass terms for fermions are,
\bea
\L &=&  \left\{
  \frac{Y_{ij}^d}{M_*} \langle H \rangle \overline{q}_i d_j
+ \frac{Y_{ij}^u}{M_*} \langle H^c \rangle \overline{q}_i u_j
\right\} \delta (y) + h.c.
\eea
(Note, the NDA estimate for the $Y$ is on the order of $6 \pi^2$
\cite{Chacko:2000hg} but we will continue to assume $Y \sim 1$ as we are
assuming a weakly coupled threshold at $M_*$).  The effective 4d masses for
the zero modes is equal to a product of the wave functions for those modes
evaluated at $y=0$,
\bea
\frac{m^d_{ij}}{v} = \frac{Y^d_{ij}}{M_*} \, \psi^0_{q_i} ( 0 ) \;
           \psi^0_{d_j} ( 0 ) & , &
\frac{m^u_{ij}}{v} = \frac{Y^u_{ij}}{M_*} \, \psi^0_{q_i} ( 0 ) \;
           \psi^0_{u_j} ( 0 ) .
\eea
In an abbreviated notation in which we write the zero mode wave function
at $y=0$ as the field itself, $i = \psi_i (0)$, we thus have the following
4d mass matrices,
\bea
\frac{m^u}{v} \sim \left(
\begin{array}{rrr}
 q_1 u_1 & q_1 u_2 & q_1 u_3 \\
 q_2 u_1 & q_2 u_2 & q_2 u_3 \\
 q_3 u_1 & q_3 u_2 & q_3 u_3 \\
\end{array}
\right) \, , \qquad
\frac{m^d}{v} \sim \left(
\begin{array}{ccc}
 q_1 d_1 & q_1 d_2 & q_1 d_3 \\
 q_2 d_1 & q_2 d_2 & q_2 d_3 \\
 q_3 d_1 & q_3 d_2 & q_3 d_3 \\
\end{array}
\right) ,
\eea
where we have suppressed the underlying ($\O(1)$) 5d interactions, $Y_{ij}$,
which multiply the corresponding entry in each matrix.  The full matrices
are thus generically of rank 3.
Assuming there is a significant hierarchy as one moves along the rows and
columns, this implies the simple relation between the three Cabibbo elements,
$V_{ub} \sim V_{us} V_{cb}$.  A further implication
is that the contributions from the up and down sectors to the CKM
matrix will be about equal in magnitude, in contrast to flavor symmetry
models.  The matrices are full in the sense that there are no negligible
entries, so for general complex $Y^u$ and $Y^d$, we should be able to realize
$CP$ violation to the extent required by measurements of $\epsilon_K$
\cite{Groom:2000in}.

\FIGURE[t]{
\epsfysize=4.0in
\centerline{\epsfbox{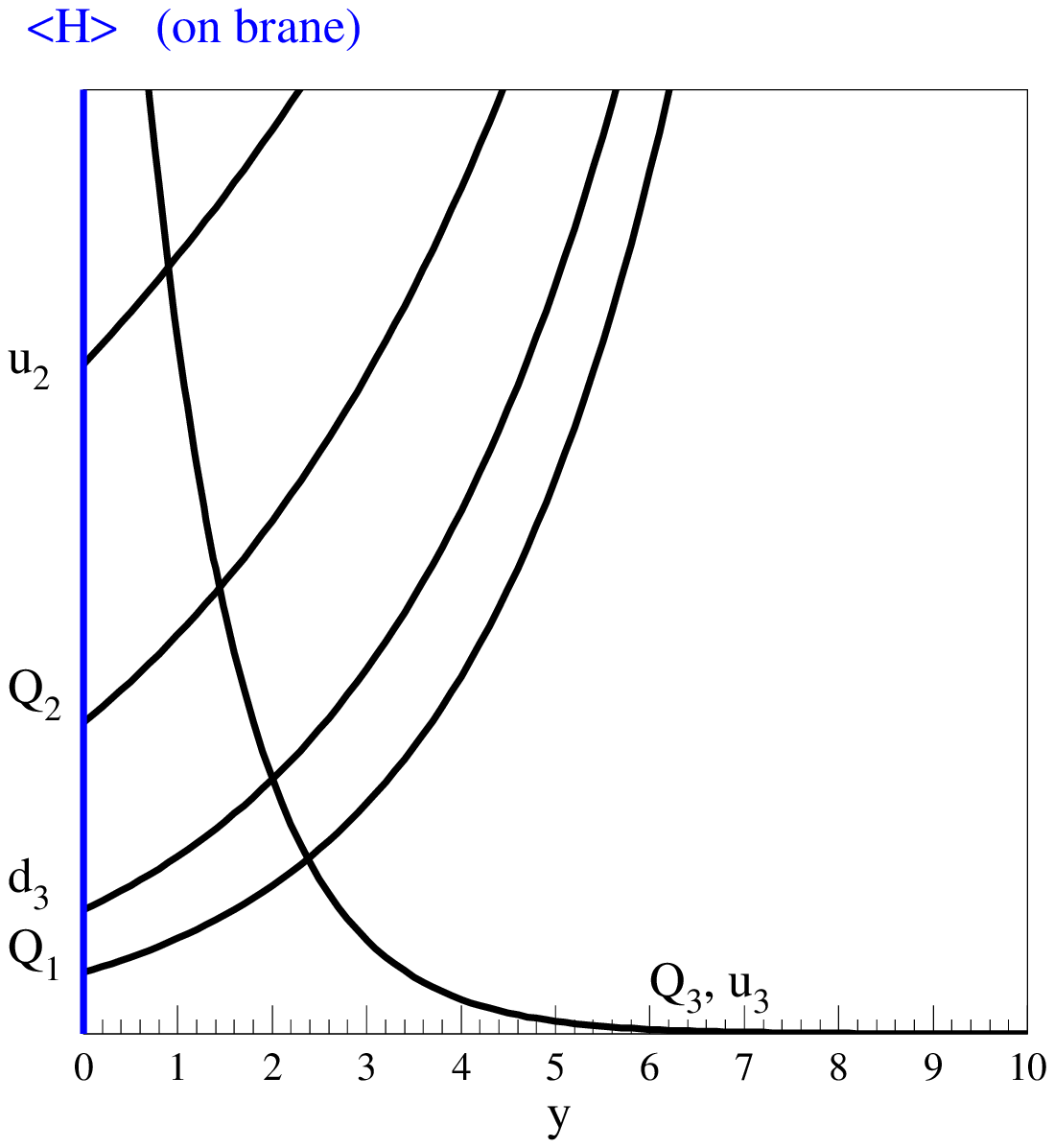}}
\caption{Zero mode profiles for some of the quarks,
for the model with the Higgs VEV localized at $y=0$.}
\label{fig:model2}}

We realize the large top Yukawa coupling by localizing $q_3$ and $u_3$
at the Higgs boundary by choosing $f_{q_3} \sim f_{u_3} \sim 1$, which results
in $y^u_{33} \sim 1$.  Note that we obtain the correct top mass without
fine-tuning simply by requiring that one quark doublet and one (up-type) quark
singlet are localized on the Higgs boundary.  This is in contrast to the
AS model where a $q$ and $u$ must be placed very close to one another.
We localize the zero modes for all of the other fermions at $y=L/2$,
and adjust the $\alpha_i$ in order to generate the observed Yukawas.
For $M_* L = 10$, this results in widths ranging from about $1/3$
(for $u_2$) to $1$ (for $u_1$), and forces us to invoke
5d Yukawa couplings ranging from about $1/3$ to $1$.
The resulting profiles for some of the zero modes are shown
in Figure~\ref{fig:model2}.

Once again, we introduce three lepton doublets, charged singlets, and
neutral singlets into the bulk, coupled to $\phi$.  We continue to assume
right-handed neutrino masses on the order of 100 TeV.
The 5d mass terms are,
\bea
\L &=&  \left\{
  \frac{Y_{ij}^e}{M_*} \; \langle H \rangle \; \overline{l}_i \, e_j
+ \frac{Y_{ij}^\nu}{M_*} \; \langle H^c \rangle \;
  \overline{l}_i \, \nu_j \right\} \delta (y)
+ M^\nu_{R ij} \; \nu^c_i \, \nu_j + h.c.
\eea
Moving to the Kaluza-Klein description, the zero modes for the
left-handed neutrinos have Dirac masses with the entire tower
of right-handed neutrino modes.  The spacing in this tower will
not be the compactification scale $1/L$ but characteristic of the
width of the localized wave function.  The contributions to the
low energy neutrino masses will differ from those estimated below
(where we only take into account zero modes) by coefficients of order
unity, which is to the accuracy we are currently working.

The Dirac masses for the charged and neutral leptons are again
proportional to wave functions evaluated at $y=0$,
\bea
\frac{m^e_{ij}}{v} = \frac{Y^e_{ij}}{M_*} \, \psi^0_{l_i} ( 0 ) \;
           \psi^0_{e_j} ( 0 ) & , &
\frac{m^\nu_{ij}}{v} = \frac{Y^\nu_{ij}}{M_*} \, \psi^0_{l_i} ( 0 ) \;
           \psi^0_{\nu_j} ( 0 ) ,
\eea
which for the charged leptons may simply be diagonalized.
Once again, we integrate out the heavy singlet neutrinos, resulting
in an effective Majorana mass matrix for the left-handed neutrinos,
\bea
\label{eq:MnuL}
M^\nu_L & = & {m^\nu} \: \frac{1}{M^{\nu}_R} \:
{m^{\nu \dagger}} .
\eea

We choose the couplings to the localizer
such that $e_3$ is localized around the Higgs VEV, and all of the other
leptons are localized around $y=L/2$.
For $M_* L = 10$, we can realize the small
mixing angle solution outlined in Section~\ref{sec:bulkhiggs} by
choosing widths ranging from $1/2$ (for $e_2$) to $2$ (for $\nu_2$)
and invoking 5d Yukawa interactions ranging from about $1/3$ to $3$.
Again, it proves somewhat difficult to realize the large mixing angle
solution.

\subsection{Constraints and the Hierarchy Solution}

Theories in which fermions live at different locations in an extra
dimensions are subject to constraints from flavor
and $CP$ violation arising from the higher KK modes of the
gauge fields \cite{Delgado:2000sv}, whose interactions depend on the
location and shape of the fermion wave function.  After performing
the CKM rotation
into the quark mass basis, this results in flavor-changing neutral
currents (FCNC's) at tree level.  While these interactions
are suppressed by the compactification scale $M_c$ that
sets the gauge boson KK masses, they may still be competitive with
the SM predictions, which occur only through loops.

For simplicity we consider only the
gluon KK modes, as they have the strongest couplings of the SM gauge
fields,
and flavor mixing in the first two generation down-type
quarks\footnote{Similar FCNC bounds on extra-dimensional lepton
flavor models were considered in \cite{Barenboim:2001wy}.}, 
which is expected to result in the strongest constraints.
The wave functions for the $n>0$ KK gluons are
$\psi_A^n (y) \sim {\rm cos} [ 2 \pi n y / L ]$ with
corresponding masses $M_n = 2 \pi n / L = 2 \pi n M_c$.  The
interaction between the $n$th KK gluon ($G^{(n)}_\mu$)
and the strange and down quarks are,
\bea
\L &=& -\sqrt{2} g_S G^{(n)}_\mu
\left( \overline{d} \: \overline{s} \right) \:
\gamma^\mu \, T^a \: \left[ D_L P_L + D_R P_R \right] \:
\left(
\begin{array}{c}
d \\
s
\end{array}
\right) ,
\eea
where the matrices $D_{L(R)}$ are defined by
\bea
D_{L} =
L_d
\left(
\begin{array}{cc}
c_1^{L (n)} & 0 \\[0.2cm]
0 &  c_2^{L (n)} \\
\end{array}
\right)
L^\dagger_d \:
&,~~~~~ &
D_{R} =
R_d
\left(
\begin{array}{cc}
c_1^{R (n)} & 0 \\[0.2cm]
0 &  c_2^{R (n)} \\
\end{array}
\right)
R^\dagger_d \:
,
\eea
a product of the left- (right-) handed down quark rotations $L_d$
($R_d$) from interaction to mass basis, and the couplings of the $n$th
KK gluon to the left- (right-) handed quark zero modes,
\bea
c^{L (n)}_i &=& \int_0^{L/2} dy \;
\cos \left[ \frac{2 \pi n}{L} y \right]
\left| \psi^0 (y) \right|^2 \; .
\eea

In \cite{Delgado:2000sv} the left-left current
contribution of the $\Delta S = 2$ portion
of these interactions to $\Delta m_K$ and $| \epsilon_K |$ was
considered. The requirement that the extra dimensional contribution
is no larger than the experimentally determined values \cite{Buchalla:1996vs}
yields the constraints\footnote{Note that our definition of $M_c$
differs by a factor of $2 \pi$ from that of \cite{Delgado:2000sv}.  Our results
for the left-left current constraints
are consistent with \cite{Delgado:2000sv} to within about 10\%,
well within the theoretical uncertainties.},
\bea
\label{eq:mkconstraint}
M_c & \gtrsim & 160 \ {\rm TeV}
\sqrt{\sum^{n^*}_{n=1} \frac{{\rm Re}
\left[ D^{2}_{L\{12\}} + D^{2}_{R\{12\}} + 14.8 D_{L\{12\}} D_{R\{12\}}
\right]}{n^2}} \, ,
\\
\label{eq:ekconstraint}
M_c & \gtrsim & 2800 \ {\rm TeV}
\sqrt{\sum^{n^*}_{n=1} \frac{{\rm Im}
\left[ D^{2}_{L\{12\}} + D^{2}_{R\{12\}} + 14.8 D_{L\{12\}} D_{R\{12\}}
\right]}{n^2}} \, ,
\eea
where we have used the vacuum insertion approximation
and the factor of 14.8 accounts for the difference in the hadronic
matrix element for a left-right as opposed to a left-left
(or right-right) current operator \cite{Beall:1982ze} as well as a relative
factor of two in the effective Hamiltonian.  The sum over KK modes is
explicitly cut-off at $n^* \sim M_* L / (2 \pi)$ to avoid counting the
modes with mass greater than $M_*$.  In fact for all of the models
we will consider there is very little sensitivity to $n_*$ because of the
$1/n^2$ suppression in the sum, as well as an additional suppression
because the high frequency modes tend to average to zero over the
quark wave functions.

Armed with the model-independent constraints Eqs.(\ref{eq:mkconstraint})
and (\ref{eq:ekconstraint}), we can now derive constraints on specific
models of large extra dimensions.  One can derive analytic expressions for
the $c^n$ constants for all of the models we have considered, but as the
expressions are somewhat unwieldy and not very illuminating, instead we
prefer to quote the resulting bounds.  The limits on $M_c$
are presented in Table~\ref{fbtab} for the three models described above.
For reference, we also show the expected relationship between the
compactification and fundamental (assumed to be related to the wave
function width) scales.
We note that our bounds on the AS model are a factor of about 50 more
stringent than those derived in \cite{Delgado:2000sv} which used a different
definition of the AS model, and included only the left-left flavor
violating currents.  In considering the bounds from $| \epsilon_K |$,
one should keep in mind that the
the AS solution contains approximate zeros in the down quark
mass matrix which would allow one to approximately rotate all of the
$CP$-violating
phases away.  This feature could allow us to interpret the bounds from
$| \epsilon_K |$, as a {\em prediction} of that model for $M_c$, as
this extra-dimensional contribution may be able to explain the experimental
measurements, though of course this would be a coincidence.

\TABLE{
\label{fbtab}
\begin{tabular}{lccc}
Quantity     & AS Model            & Higgs in the Bulk  & Higgs on a Brane\\
\hline
$M_c \;{\rm from}\; \Delta m_K$  & 120 TeV       & 5 TeV  & 13 TeV          \\
$M_c \;{\rm from}\; \epsilon_K$ & 2100 TeV       & 80 TeV & 230 TeV         \\
$M_*$           & $50 \times M_c$ & $10-20 \times M_c$ &
$10-20 \times M_c$ \\
\end{tabular}
\caption{Limits from Kaon measurements on the three models described
in the text.}}

As the table shows, the models we have constructed in which the quarks
and leptons live on one or the other of the orbifold fixed points have
significantly weaker bounds on $M_c$ than the AS model.  This can be
understood largely from the fact that in the orbifold models the first
and second generation down-type quarks are localized about the same point,
with differences in masses and mixings arising from the different widths of
the wave functions, whereas in AS the quarks are localized at different points
and thus for the lower KK modes of the gluon (which dominate the sum in
Eqs.(\ref{eq:mkconstraint}-\ref{eq:ekconstraint}) the couplings to the
two quarks are more equal, and thus the flavor-violation less pronounced.
Furthermore, the AS model requires the fundamental scale be considerably
higher than the other models, because it must actually space the multiple
fermions away from each other to get small masses {\em and} mixing angles.

If one makes the extra dimension in AS a bit larger, one can incorporate
their solution to the problem of proton decay via higher dimension operators.
Their solution, separating quarks and leptons in the bulk, is not easily
adapted into our framework and thus we require further ingredients
(for example, imposing additional gauge symmetries could forbid the
dangerous operators) to be consistent with proton decay constraints.

\section{Supersymmetric Models and $y$-dependent Flat Directions}
\label{flatd}
\indent \indent
We now promote the above models to supersymmetrized versions using the
superfield notation of \cite{Arkani-Hamed:2001tb} for dimensions
greater than four (which we review below).  We find that using odd
mass terms is enough to generate the complete Yukawa hierarchy.
We also find scalar VEV profiles which preserve ${\cal N}=1$
supersymmetry and can be used to localize chiral superfields.  While
the restrictive nature of 5d supersymmetry makes it difficult to
construct realistic models, we do outline a working model in this context
compactified on an orbifold.
Finally, we present a model compactified
on $S^1$ where chirality is simply introduced by introducing a brane
containing chiral matter.  Vector-like fields in the bulk mix with the
chiral fields on the brane and a scalar VEV ``projects'' the massless
chiral fermions into the bulk.

\subsection{Superspace for five-dimensional supersymmetry}

Using the notation of \cite{Arkani-Hamed:2001tb}, we formulate a 5d
supersymmetric gauge theory in the language of ${\cal N}=1$ 4d superfields.
This allows us to use the powerful superfield machinery to analyze the
conditions under which one supersymmetry is preserved in the 4d effective
theory.  We find $y$-dependent flat directions which then can be used as
tools for model-building.

\subsubsection{Hypermultiplets}
Chiral superfields in five dimensions come in pairs, called hypermultiplets.
A free hypermultiplet was first described in the following notation in
\cite{Arkani-Hamed:2001pv}:
\begin{equation}
\int d^4 \theta \left(H^{\dagger} H + H^{c\dagger} H^c\right)
+ \int d^2 \theta \: H^c \left(\partial_y + m\right) H + {\rm h.c.}
\end{equation}
If the fifth dimension is compactified on a ${\cal Z}_2$ orbifold, then
$H$ and $H^c$ must transform oppositely under the discrete symmetry
and thus $m = 0$.  Another option
is to give the hypermultiplet a mass which is odd under the ${\cal Z}_2$
with one chiral superfield odd and the other even.  This mass term preserves
the full 5d supersymmetry everywhere except at the boundaries, where it
preserves half.

\subsubsection{An Abelian Gauge Multiplet}
The 5d gauge sector consists of a vector superfield $V$ whose components are
the four-dimensional part of the vector gauge field $A^\mu$, the left-handed
gaugino $\lambda_L$, and an auxiliary field $D$, and a chiral superfield
$\Phi$ whose components are a complex scalar $\phi =  (\Sigma + i A_5)/\sqrt{2}$
(containing both the fifth component of the vector field $A_5$ and the real
scalar $\Sigma$), the right-handed gaugino $\lambda_R$, and a complex auxiliary
field $F$.  The 5d Lagrangian density is
given by
\begin{eqnarray}
&&\int d^4 \theta \frac{1}{g^2}
	\left(\Phi^{\dagger} \Phi
		- \sqrt{2} \left(\Phi^{\dagger} + \Phi\right)\partial_y V
        	- V\partial_y^2 V \right) \nonumber\\
&+& \int d^2 \theta \frac{1}{4 g^2} W_{\alpha}W^{\alpha} + {\rm h.c.} \, .
\label{eq:u1}
\end{eqnarray}
While this Lagrangian is only manifestly 4d Poincar\'{e} invariant, it is
in fact invariant under the full 5d Poincar\'{e} symmetry.  It is also
invariant under the 5d gauge transformations:
$V \ra V + \Lambda^\dagger + \Lambda$ and
$\Phi \ra \Phi + \sqrt{2} \partial_y \Lambda$,
as well as the full ${\cal N}=2$ supersymmetry transformations
\cite{Arkani-Hamed:2001tb}.

\subsubsection{Charged matter}
A hypermultiplet of charge $Q$ consists of two chiral superfields
$H$ and $\Ht$ with scalar components $h$ and $\htilde$, fermionic
components $\psi_h$ and $\psi_{\htilde}$, and auxiliary fields $F_H$
and $F_{\Ht}$ and the following terms in the Lagrangian:
\begin{eqnarray}
&&\int d^4 \theta \left[H^{\dagger} e^{-Q V} H + \Ht^{\dagger} e^{Q V} \Ht
        \right] \nonumber\\
&+& \int d^2 \theta
\left[\Ht \left( \partial_y + m - \frac{Q}{\sqrt{2}} \Phi \right) H \right]
	+ {\rm h.c.}
\label{eq:hyper}
\end{eqnarray}
Generalizing to more than one hypermultiplet is trivial.  For hypermultiplets
of the same charge, $m$ can be a matrix with non-trivial flavor structure.
Under gauge transformations, the hypermultiplet transforms as
$H \ra e^{ Q \Lambda } H$, and $\Ht \ra e^{ -Q \Lambda } \Ht$.

\subsubsection{Coupling to Branes/Boundaries}
One of the reasons this notation is so powerful is that it makes coupling
bulk fields to branes trivial.  For example, a superpotential coupling
of a component $H$ of an uncharged hypermultiplet to a brane at $y=0$
would look like
\begin{equation}
\label{eq:FIterm0}
\int d^2 \theta \: J H \delta(y) \, ,
\end{equation}
where $J$ is a gauge invariant operator made up of brane fields and/or
numerical constants.  A Fayet-Iliopoulos term on a brane at $y=0$ looks like
\begin{equation}
\label{eq:FIterm}
\int d^4 \theta \: 2 \, \xi \: V \: \delta(y) \, ,
\end{equation}
while adding charged fields $X,{\tilde X}$ (with charges $\pm 1$) to a
brane at a point $y=L/2$ requires
\begin{equation}
\label{eq:branefields}
\int d^4 \theta \:
\left( X^{\dagger} e^{-V} X + {\tilde X}^{\dagger} e^V {\tilde X}\right)
\delta(y - L/2) \, .
\end{equation}
When translated into component language, this notation reproduces
the results of \cite{Mirabelli:1998aj}.

\subsubsection{Flat directions}
We now have the machinery needed to look for $y$-dependent flat directions
which preserve the 4d ${\cal N}=1$ supersymmetry.  We
simply need to solve the $F$- and $D$-flat conditions.  Before we do,
we remind the reader that our fifth dimension is compact.
We are interested in both compactification on a simple circle ($S^1$),
with $-L/2 < y \leq L/2$, and on an orbifold ($S^1 / \Z_2$), with the
same range for $y$ but with $y$ and $-y$ identified.  In the latter case,
the superfields $H$ and $V$ are even under the $\Z_2$ and $\Ht$ and $\Phi$
are odd.  This has the consequence (as in the previous section) that a normal
mass term connecting $H$ with $\Ht$ is forbidden while an odd mass term
(proportional to $\epsilon(y)$) is allowed.

First let us look at the case with a U(1) vector multiplet with a
Fayet-Iliopoulos term and matter with charges $\pm Q$ on branes
(or orbifold fixed points) at $y=0$ and $y=L/2$ respectively.
The $D$-flat condition requires
\begin{eqnarray}
\label{eq:flatd1}
-D & = & \left[ 2\xi\delta(y)
	+ \frac{g \, Q}{2} \left( |{\tilde X}|^2 - |X|^2 \right)\delta(y-L/2)
	+ \partial_y \Sigma \right] = 0 \: ,
\end{eqnarray}
which is satisfied by the conditions $| {\tilde X}|^2 - |X|^2 = - 4\xi/g Q$
and $\Sigma = \Sigma_0 + \xi\epsilon(y)$.  In the case of the orbifold,
$\Sigma_0 = 0$ since $\Sigma$ is odd around the points $y=0,L/2$.
It is simply a degree of freedom which is projected out of the theory by
the orbifold \cite{Arkani-Hamed:2001tb}.

As can be seen from Eq. (\ref{eq:hyper}), $\Phi$ can play the role of the
localizer field\footnote{Localizing chiral
superfields requires a straight-forward generalization of the procedure
for localizing a fermion zero mode.  For details, see \cite{Kaplan:2000av}.}
with standard model matter (and their mirror partners)
as hypermultiplets in the bulk.  The Lagrangian contains
$\psi_{\tilde q}(D_y + Q \Sigma(y)/2)\psi_q$ which localizes the zero mode
of the
quark $\psi_q$(mirror-quark $\psi_{\tilde q}$) where $\Sigma(y)$
crosses zero with a positive (negative) slope.  In a compact space,
zero modes only exist for $\Sigma_0 = 0$.  In
the orbifold case this condition, as well as the removal of the $\tilde q$
zero mode, is guaranteed by the boundary conditions.  In the $S^1$ case this
could be guaranteed by a soft mass
for $\Sigma$ on either brane.  If $\Sigma_0$ is non-zero, the lightest mode
mass goes as $\sqrt{\xi^2 - \Sigma_0^2}\, e^{-(\xi - \Sigma_0)L/2}$ for
$\Sigma_0$ at least somewhat smaller than $\xi$.

More interesting VEV profiles can appear if we include a hypermultiplet in
the flat directions.  Using equations (\ref{eq:u1}) and (\ref{eq:hyper}),
we look for solutions to the differential equations resulting from imposing
the $F$- and $D$-flatness conditions
$D^2 = |F_H|^2 = |F_{\Ht}|^2 = |F_\Phi|^2 = 0$, where,
\begin{eqnarray}
\nonumber
-F_{\Phi}^{*} & = & -\frac{g \, Q}{\sqrt{2}} \htilde h \\
-F_{\Ht}^* & = & \left[ \partial_y + m - \frac{Q g}{\sqrt{2}} \phi \right] h
\nonumber\\
-F_{H}^* & = & \left[ -\partial_y + m - \frac{Q g}{\sqrt{2}} \phi \right]
\htilde
\nonumber\\
-D & = & \frac{g \, Q}{2} \left( | \htilde |^2 - |h|^2 \right)
	+ \partial_y \Sigma \: .
\label{eq:flatd}
\end{eqnarray}
As it turns out for the $S^1$ and $S^1 / \Z_2$ geometries, the only
solutions are $\Sigma = \Sigma_0$ and $\phi = 0$ respectively with
all other scalar fields zero.  This is because the compactification of the
extra dimension requires solutions which are periodic, and while such
solutions to Eqs.(\ref{eq:flatd}) exist, they have VEV's which
are singular at points in the extra dimension, and thus our effective theory
description of the physics may not be applicable.
In order to have nontrivial profiles valid within the
context of the effective theory, we introduce a 3-brane located at
$y = 0$ with a Fayet-Iliopoulos term (\ref{eq:FIterm}),
which modifies the $D$ term equation as in (\ref{eq:flatd1}).  This in
turn induces a discontinuity in the VEV of $\phi$ at the brane.
The $F_\Phi$ and $F_H$ equations may be satisfied by requiring
$\htilde = 0$, and the remaining two equations have solutions,
\begin{eqnarray}
h(y) & = & \frac{2 \alpha}{g \cos[\alpha(y+L/2 - L\Theta(y))]}
\nonumber \\
\Sigma (y) & = & \frac{\alpha}{g} \tan[\alpha(y+L/2 - L\Theta(y))]
	+ \frac{m}{g},
\label{eq:u1profiles}
\end{eqnarray}
where we have taken the charge $Q=1$ and explicitly
chosen a (5d) gauge to make $h(y)$ real and $A_5$ vanish.
The parameter $\alpha$ is related to the magnitude of the
Fayet-Iliopoulos term by,
\begin{equation}
\alpha = \sqrt{\frac{g \, \xi}{2 \, L}} \, ,
\end{equation}
and nonsingular VEV's in the interval $-L/2 \leq y \leq L/2$
require $\alpha<\pi/L$.

Again, the profile for $\Sigma$ acts as a ``mass function'' and will tend to
localize the right-handed components of (positively charged) hypermultiplets
about the point where the brane sits ($y=0$) and the left-handed components
at a point in the bulk where $\Sigma(y)$ crosses zero \cite{Kaplan:2000av},
with the KK tower masses given as eigenvalues of the supersymmetric
QM Hamiltonians,
$-\partial^2_y \mp g / \sqrt{2} (\partial_y \Sigma(y)) + g^2 / 2 \Sigma^2(y)$ .
As an example in order to divine some
general features, we consider the case where
$m=0$ (which would be enforced, for example, by orbifold boundary
conditions) and will allow for the lightest modes in the
KK decomposition
to have zero mass.  Our analysis is further simplified when
$\alpha$ is small, which allows us to expand the profile for $\Sigma$ as,
\bea
\Sigma(y) &=& \frac{\alpha^2}{g} \left[ y + L/2 - L \Theta(y) \right]
\, .
\eea
The zero mass solutions for a hypermultiplet
(containing chiral multiplets $\tilde{\Psi}$ and $\Psi$) of charge $Q$ are,
\bea
\psi^0_{\pm}(y) &=& {\cal N}_{\pm}
{\rm Exp} \left[
\pm Q \alpha^2 \left( \frac{1}{2} y^2 + \frac{L}{2} y - L y \Theta(y)
\right) \right] ,
\eea
where ${\cal N}_{\pm}$ are chosen to correctly normalize the kinetic
terms.  In the limit of large $L$, these solutions look
increasingly like an exponential centered at $y=0$ and a Gaussian
centered at $y= L/2$, which is understandable because in that limit
the corresponding Hamiltonians look like a $\delta$-function potential
at $y=0$ and a simple harmonic oscillator at $y=L/2$, each
surrounded by large ``potential barriers'' that discourage the wave
functions of the low mass modes from spreading.

If we now allow non-zero $\Delta m = m_H - m$,
the situation changes in two important ways.
The zero-crossing of the linear term in $\Sigma$ will shift,
which will translate the center of the part of the
Hamiltonian which looked like a harmonic oscillator (if $\Delta m$ is large
enough, the zero crossing may in fact disappear altogether).  More importantly,
the two fields which formerly corresponded to the right- and left-handed
zero modes will now marry one another with some non-zero mass of
${\cal O}(\Delta m)$.  However, provided $L$ is large
(and thus the potential barrier between the two localizing potentials in the
corresponding Schr\"odinger problem is also large), the profiles for this pair
of light modes remain localized as they were for the $m=0$ case.  Thus, the
lightest modes of the KK spectrum are a pair of chiral superfields
($\tilde{H}^0$ and $H^0$) localized at $y=0$ (with approximately
exponential profile) and $y= L/2 - \Delta m \, Q / \alpha^2$
(with approximately Gaussian profile), respectively.

%

\subsection{Flavor from an Odd Mass Term}
\label{sec:susyflavor}

The simplest and perhaps most attractive model of the sort we are discussing
requires masses for hypermultiplets which are odd under the ${\cal Z}_2$
of the orbifold\footnote{We thank Andrea Romanino, who was the first to point
out this possibility to us.}.  The hypermultiplets are the SM quarks and
leptons and their 5d chiral partners.  The odd mass localizes the chiral
zero modes of the matter fields at one of the orbifold fixed points, depending
on the sign of the mass-term step function.

The model reproduces (a supersymmetrized version of) the model in
Section \ref{sec:localhiggs} with the scalar profile idealized to a
step function.
The Higgses are now chiral superfields and can
be localized in the same way as the matter fields.
The Yukawa interactions are forbidden by $\N =2$ supersymmetry,
but may be explicitly introduced on the branes.
Thus the most successful model of the previous section can be put into a
supersymmetric context and thereby decouple the gauge hierarchy problem from
the generation of Yukawa suppression.  In fact, because generic supersymmetric theories
have two Higgs doublets, one to generate up-type quark
(and, if relevant, Dirac neutrino) masses, and the other to generate down-type
quark and charged lepton masses, we have an additional freedom in constructing
flavor in a supersymmetric context in the choice of ratio of the Higgs
VEV's, $\tan \beta = v_u / v_d$.  This allows one, for example, to
partially or completely generate the hierarchy between the top and bottom
masses by the choice of $\tan \beta$, and allows some more flexibility
in generating realistic flavor.

The next question is whether or not one can promote the above mass
term to a field such that one can produce successful models of flavor
with localized fermions zero modes by the $y$-dependent profiles
described above.
The short answer is no.  Coupling a hypermultiplet to the $\Sigma$
field in the superpotential requires the hypermultiplet to be charged
under a gauge symmetry.  Taking that gauge symmetry to be U(1), the
couplings of matter fields to $\Sigma$, and thus the width of their
wave functions, are proportional to the charge of the hypermultiplet
in question.

If one wants different widths for different
generations, the fields must have different charges.  However, this
forbids most or all of the Yukawa couplings in the 5d theory.  One could
choose charges $Q$ such that $Q_{q_i} + Q_{u_i} + Q_{h_u} = 0$ and
$Q_{q_i} + Q_{d_i} + Q_{h_d} = 0$, where $i=1,2,3$ is the generation
index, and so at best one can get the right mass hierarchies in both the
up and down sectors (with the $\mu$ term forbidden by the U(1) symmetry).
However, the Yukawa matrices will already be diagonal and thus the
CKM matrix is the identity matrix.
One can remedy this situation by noticing that the boundary
fields required to produce the $\Sigma$ profile break the U(1) gauge
symmetry spontaneously.
This field can be used to produce non-renormalizable operators
which could allow mixing terms once the field's VEV is inserted.
The result is a hybrid extra-dimensional/Froggatt-Nielsen mechanism
for fermion masses.  While this idea seems workable, the resulting
models are more in the Froggatt-Nielsen spirit than an extra-dimensional
one, so we will not pursue them here.

\subsection{Compactifying on $S^1$}
\label{sec:s1}

An alternative to the orbifold is to work with the extra dimension
compactified on $S^1$, and introduce chiral matter explicitly on a 3-brane.
To illustrate how this works,
let us consider a  bulk hypermultiplet containing chiral multiplets
$\Psi$ and $\Psi^c$.  We include a 3-brane at $y=0$ on which
lives a chiral superfield $\eta$, and include a brane-coupling between
$\eta$ and $\Psi^c$,
\bea
\int dy \: \int d^2 \theta \: \Psi^c (\partial_y + m) \Psi
+ M \: \eta \, \Psi^c \; \delta( y ) \: .
\eea
Without the orbifold, $\Psi$ is allowed an ordinary mass $m$.
Ignoring the brane coupling for the moment, we consider the case in
which $\Psi$ is charged with charge $Q$ under a bulk U(1) whose $\Sigma$ is
given the profile of Eq.~(\ref{eq:u1profiles}). Its lightest KK mode $\Psi^0$
will tend to localize around the zero crossing of the function
$(Q \times \langle \Sigma \rangle(y) - m)$.  Of course, it will have some
non-zero mass $m_0$ with $\Psi^{c\,0}$, which will tend to localize
around the brane (the locations can be reversed by adjusting the sign
of $Q$ and/or $m$).  If we now turn on the brane coupling,
the net result will be one massless field and one
with mass $\sqrt{ M_0^2 + m_0^2 }$ (with $M_0$ given by the overlap of
the $\Psi^{c\,0}$ wave function with the brane), each of which is a
mixture of the bulk light mode $\Psi_0$ and the brane field $\eta$.
The composition of the zero mass field will be
\bea
-\frac{M_0}{\sqrt{M_0^2 + m_0^2}} \Psi_0 + \frac{m_0}{\sqrt{M_0^2 + m_0^2}}
\eta \; ,
\eea
indicating that provided $M_0 \gg m_0$, we have essentially recovered
a chiral field in the bulk (though with some small component living
on the brane).  The chiral fields on the brane have been ``projected''
from the brane into the bulk by appropriate mixing with bulk fields.

We can use this tool to avoid the problems of the $S^1 / \Z_2$ models in the
previous section.  We introduce a brane containing the entire MSSM chiral
superfield sector, with brane couplings to an entire MSSM hypermultiplet
sector in the bulk\footnote{For producing the right masses in the charged
sector, it is in fact not necessary to put an entire
MSSM hypermultiplet sector in the bulk.  Simply a set of (what makes up)
10's and their conjugates will do.}.

In order to allow for Higgs couplings, we assign each
generation the same charges for a given type of field, for
example:
$Q_q = +1/2$, $Q_{u^c} = +1$, $Q_{d^c} = -2$, $Q_l = +1/2$, $Q_{e^c} = -2$,
$Q_{n^c} = +1$, $Q_{H_u} = -3/2$ and $Q_{H_d} = +3/2$.
These charges allow inter-generational couplings to the Higgses (on the branes),
and results in the bulk light modes for the $q$, $u^c$, $l$, $n^c$, and $H_d$ fields
living at various points in the bulk (with positions determined by the corresponding
hypermultiplet masses) and the $e^c$, $d^c$, and $H_u$ fields all living on
the brane.  The right-handed neutrino masses are now forbidden
by the U(1) symmetry, but could be generated by a
non-renormalizable superpotential term such as $H H n^c n^c$.
By appropriately choosing the bulk masses, we may adjust
the overlaps of the left-handed fields with the right-handed fields and
Higgses, and thus realize viable flavor.  This mechanism has something
in common with both the AS mechanism in that one sees suppression from
right- and left-handed fields overlapping, and also some features of
suppression due to the overlap with the Higgs present in the models
of Section~\ref{sec:localhiggs} and \cite{Kaplan:2000av}.

\section{Supersymmetry Breaking}
\label{susyb}
\indent \indent
Having successfully constructed supersymmetric theories in which
flavor is generated by an extra-dimensional mechanism, it is important
to also consider how supersymmetry is broken.  A generic supersymmetry-breaking
mechanism could lead to off-diagonal entries in the sfermion mass matrices.
The simplest way to avoid this supersymmetric flavor problem is to break
supersymmetry in such as way as to guarantee that all sfermions of the
same charge have approximately degenerate masses.  This insures that after
the rotation from gauge to mass eigenstates required to diagonalize the
fermion masses, the sfermion masses remain diagonal.  Since we have already
introduced an extra dimension, we briefly consider two extra-dimensional
supersymmetry-breaking mechanisms: Scherk-Schwarz breaking \cite{Scherk:1979ta}
by twisted boundary conditions (as realized in \cite{Barbieri:2001yz}), and
gaugino-mediation \cite{Kaplan:2000ac,Chacko:2000mi}.

\subsection{Scherk-Schwarz Breaking}

Any of the flavor models of the previous section can
incorporate supersymmetry breaking by modifying the
orbifold boundary conditions on the components of the superfields such
that masses of the superpartner zero-modes are lifted to weak scale
values.  The model, and discussion, follows closely the one proposed
in \cite{Barbieri:2001yz}.  As before,
we break $\N=2$ down to $\N=1$ by requiring that
under the identification $y \leftrightarrow -y$ the superfields transform as,
\bea
\left( \begin{array}{r}
V \\ \Phi \end{array} \right) (x^\mu, -y)
&=& \left( \begin{array}{r} V \\ -\Phi \end{array} \right) (x^\mu, y), \\
\left( \begin{array}{l} \Psi \\ {\Psi}^c \end{array} \right)
(x^\mu, -y) &=&
\left( \begin{array}{c} ~ \Psi \\ -{\Psi}^c \end{array} \right)
(x^\mu, y) .
\eea
where $\Psi$ and $\Psi^c$ together form one of the
matter hypermultiplets.
Under $y \leftrightarrow y + 2 \pi R$, the two gauginos
and two sfermions are twisted into each other by an element of
the $SU(2)_R$ symmetry of the 5d theory,
\bea
\left( \begin{array}{r} \lambda_L \\ \lambda_R \end{array} \right)
(x^\mu, y + 2 \pi R) &=&  e^{ -i 2 \pi \alpha \sigma_2 }
\left( \begin{array}{r} \lambda_L \\ \lambda_R \end{array} \right)
(x^\mu, y), \\
\left( \begin{array}{l} \widetilde{f} \\ \widetilde{f}^c \end{array} \right)
(x^\mu, y + 2 \pi R) &=& e^{ -i 2 \pi \alpha \sigma_2 }
\left( \begin{array}{l} \widetilde{f} \\ \widetilde{f}^c \end{array} \right)
(x^\mu, y)
\eea
where $\sigma_2$ is the Pauli matrix, and $\alpha$ is a dimensionless
parameter specifying the amount of twisting.
The vectors, gauge scalars, and fermions are untwisted, and will thus
remain as zero modes in the low energy theory.

The additional boundary conditions on the fields modify the KK expansion
for the gaugino modes to,
\bea
\left( \begin{array}{r} \lambda_L \\ \lambda_R \end{array} \right)
(x^\mu, y) &=& \sum_{n}
e^{ -i \alpha y / R \sigma_2 }
\left( \begin{array}{r} \lambda^n_L \; {\rm cos} \left[ n y / R \right]
\\ \lambda^n_R \; {\rm sin} \left[ n y / R \right] \end{array} \right) ,
\eea
which, substituted into the 5d action \ref{eq:u1} and integrating
over $y$ results in universal masses $\alpha / R$ for the
gaugino zero modes.  Assuming a compactification scale close to the
GUT scale, this requires $\alpha \sim 10^{-13}$ in order to have
gaugino masses at the weak scale.

The scalar masses are slightly more subtle.  First, we note that the
matter fermions have untwisted boundary conditions, and so are localized
exactly as before, with wave functions $F^n(y)$ for the fermions (including
a zero mode) and wave functions $G^n(y)$ for the mirror fermions.
In this basis the KK expansion for the sfermions is
\bea
\left( \begin{array}{l} \widetilde{f} \\ \widetilde{f}^c \end{array} \right)
(x^\mu, y) &=& \sum_{n}
e^{ -i \alpha y / R \sigma_2 }
\left( \begin{array}{l} \widetilde{f}_n \; F^n(y)
\\ \widetilde{f}_n^c \; G^n(y) \end{array} \right) .
\eea
Inserting this expansion into the 5d the kinetic term produces universal
sfermion masses $\alpha^2 / R^2$.  Flavor-dependent corrections to the wave
functions and masses will appear at order $\alpha / R$ and thus are 
negligibly small.

This model manages to generate the correct fermion spectrum while 
avoiding supersymmetric flavor problems.  This is in contrast to a 
Froggatt-Nielsen type of mechanism which, if the flavor-breaking scale 
is at least somewhat below the compactification scale, will produce 
flavor-violation in the scalar sector through renormalization group running.  
Unfortunately, the mechanism of flavor generation has virtually no 
impact on the superpartner spectrum and thus would be difficult to study 
experimentally at energies far below the compactification scale.

\subsection{Gaugino Mediation}

\FIGURE[t]{
\epsfysize=2.0in
\centerline{\epsfbox{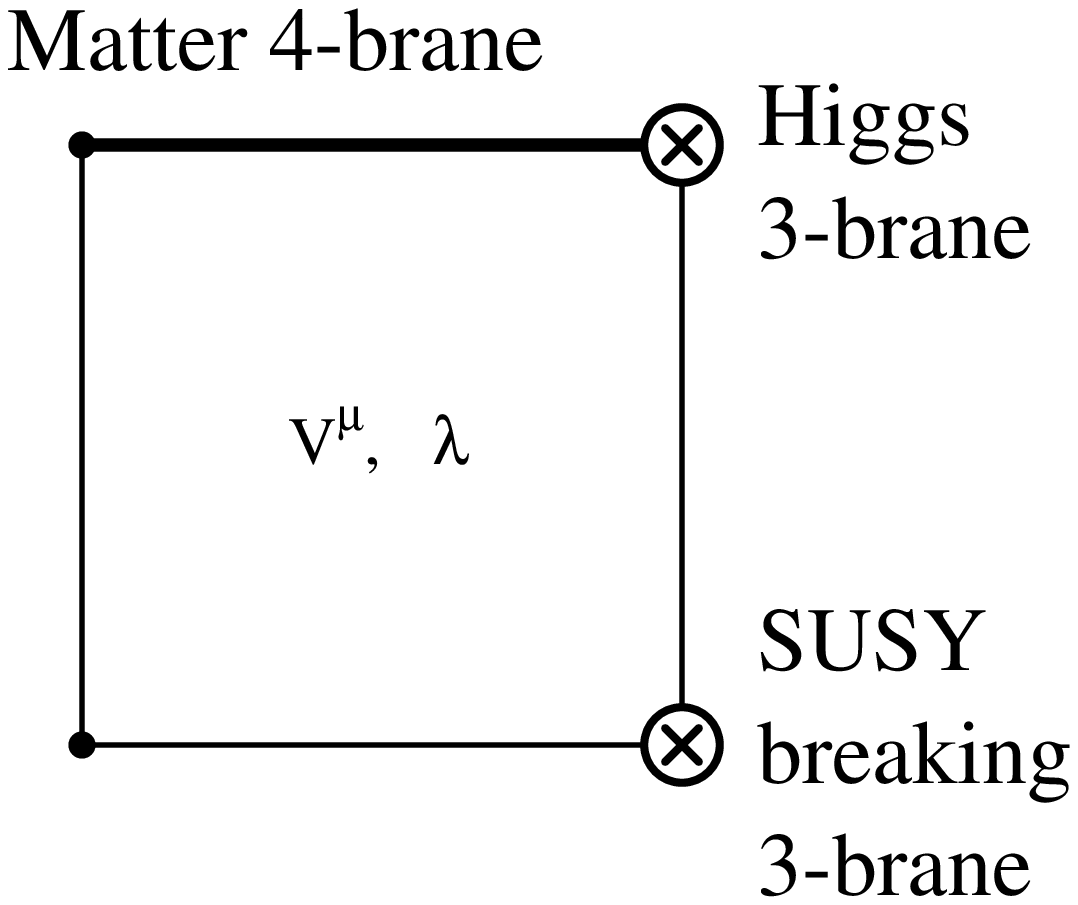}}
\caption{Schematic of the 6d model, indicating the locations
of the Higgs and supersymmetry-breaking 3-branes, the matter 4-brane, 
and the gauge and gaugino fields in the bulk.}
\label{fig:6d}}

In order to simply imbed the model of Section~\ref{sec:susyflavor} in a model
of gaugino-mediation, we consider a theory with {\em two} extra
dimensions compactified as $T^2 / \Z_2$, a 2-torus with
two points mapped into each other by a $\pi$ rotation
in the plane of the compact dimensions identified.
The coordinates in the extra
dimensions can be expressed as a 2-vector $\vec{y} = (y_1, y_2)$,
with the physical space lying inside the rectangle bounded by
the four orbifold fixed points at $(0,0)$, $(\pi R_5, 0)$,
$(0,\pi R_6)$, and $(\pi R_5, \pi R_6)$ \cite{Hall:2001xr}.
For simplicity, we consider the case where the two compact
dimensions are orthogonal, and both radii are equal,
$L = \pi R_5 = \pi R_6$.  The gauge
fields live in the entire 6d bulk, with the quarks and leptons confined
to a 4-brane stretching between two of the orbifold fixed points
(with zero modes localized along the small brane direction
in order to produce flavor as in Section~\ref{sec:susyflavor}), and the
Higgses live in a 3-brane located at one of these two points.  Supersymmetry
is broken at one of the two-fixed points outside of the matter-brane.
The situation is shown schematically in Figure~\ref{fig:6d}.

If we parameterize the supersymmetry-breaking by a chiral superfield $X$ whose
auxiliary component $F_X$ has a non-vanishing VEV, gauginos acquire a mass
at tree-level from effective interactions such as,
\bea
\label{eq:gauginomass}
\int d^6 x \, \int d^2 \theta \;
\frac{\lambda_X}{M_*^3} X \: W^\alpha \, W_\alpha \: \delta ( \vec{y} ) ,
\eea
where $\lambda_X$ is a dimensionless coupling of order unity, and
$\vec{y} = (y_1, y_2)$ are the coordinates in the extra dimensions.  This
results in a mass for the zero-mode gaugino of
$\lambda_X \langle F_X \rangle / M_*^3 L^2$,
suppressed by the volume of the extra dimension.

The sfermions
are prevented from getting masses (or $A$ terms) directly from
$X$ by 6d locality, and must instead acquire masses at one loop
from the gauginos through Feynman graphs such as that shown in
Figure~\ref{fig:gauginomed}.
Below the compactification scale,
the only relevant contribution from this graph has the gaugino zero-mode
in the loop (which in fact corresponds to the usual renormalization
group evolution of the sfermion mass induced by the gauginos in 4d).
Since the zero-mode has flavor-blind couplings to the sfermions, this
results in universal sfermion masses, as desired.  However, above the
compactification scale the higher KK modes of the gaugino will also
contribute to the sfermion masses, and since they have wave functions
which vary across the extra dimension, they couple flavor-diagonally,
but not flavor-independently.  This is potentially a problem, because
after the rotation from the gauge to mass basis, the sfermions will, in
general, pick up off-diagonal entries proportional to the mas${\rm s}^2$
differences multiplied by rotation angles.

\FIGURE[t]{
\epsfysize=1.0in
\centerline{\epsfbox{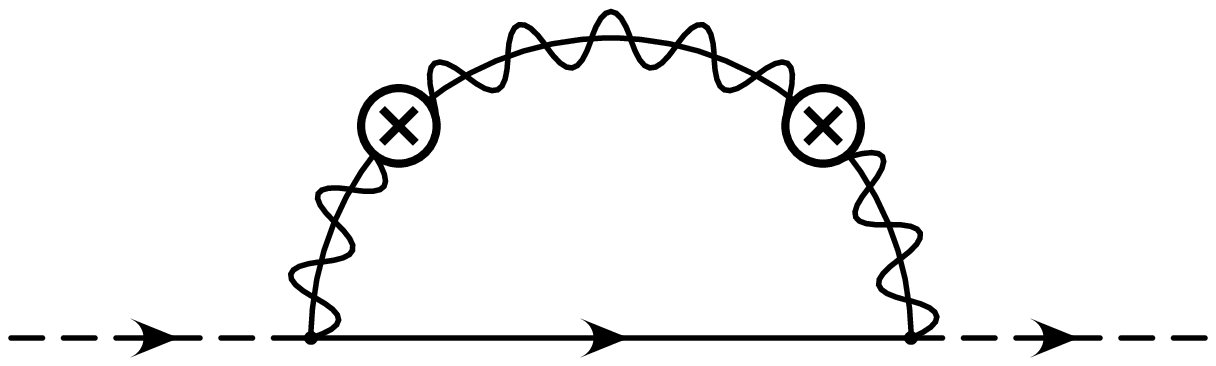}}
\caption{Loop diagram showing how gauginos carry supersymmetry-breaking information
to the sfermions.  The $\otimes$'s represent insertions of the
operator in Eq.~(\protect{\ref{eq:gauginomass}}).}
\label{fig:gauginomed}}

These contributions may be estimated by expressing the 6d gaugino propagator
$\P[q; \vec{a}, \vec{b}]$
in mixed position and momentum space (see, for example \cite{Kaplan:2000av}),
and evaluating the 6d effective action at one loop, summing over all
of the gaugino KK modes in the loop\footnote{In six dimensions, this
introduces some dependence on how the sum over KK modes is cut-off;
we have adopted a hard cut-off at $M_*$.} and identifying the
term relevant for sfermion masses,
\bea
\Gamma_6 \left[\widetilde{f}^*,\widetilde{f}\right] &=&
\int d^4x \, dy_1 \, dy_2 \:
\widetilde{f}^*(x^\mu, y_1)
\, \widetilde{f}(x^\mu, y_2) \, M^2 (y_1, y_2) ,
\eea
where $y_{1(2)}$ are positions along the matter 4-brane, and we have
explicitly used the fact that a 4d mass term must be evaluated for both
fields at the same 4d space-time point, and that the (s)fermions are confined
to a 5d subspace.  The coefficient $M^2(y_1, y_2)$ is,
\bea
M^2(y_1, y_2) &=& \frac{\alpha}{4 \pi} M^2_{1/2}
\int d^4 q {\rm Tr} \left[ \frac{\qslash}{q^2} \P[q; (L,y_1), \vec{0}]
\P[q; \vec{0}, \vec{0}] \P[q; \vec{0}, (L, y_2)] \right] ,
\eea
where $\alpha$ is defined in terms of the 4d gauge coupling, and we have
dropped Casimirs and factors of 2.  One may then compactify
down to four dimensions, inserting the wave functions for the (s)fermions
and determining the effective mass of the sfermion ``zero-mode'' at the
compactification scale,
\bea
\widetilde{m}^2 &=& \int dy_1 \, dy_2 \; \psi^{0 *}_{\widetilde{f}}(y_1)
\; \psi^{0}_{\widetilde{f}}(y_2) \: M^2(y_1, y_2) .
\eea
The weak scale superpartner masses are then
obtained by rotating to the quark mass basis, and applying the usual
4d renormalization evolution from the compactification scale to the weak scale.

For simplicity we assume $M_c \sim M_{GUT}$, with the gaugino masses
given by a single parameter $M_{1/2}$.  The boundary conditions
for the scalar mas${\rm s}^2$ matrices at $M_c$ for the specific
localized Higgs model described in Section~\ref{sec:susyflavor} results in,
\begin{alignat}{3}
\widetilde{m}^2_{q_1} & \sim 0.024 \; M_{1/2}^2, \qquad
\widetilde{m}^2_{u_1} & \sim 0.023 \; M_{1/2}^2, \qquad
\widetilde{m}^2_{d_1} & \sim 0.023 \; M_{1/2}^2, \nonumber \\
\widetilde{m}^2_{q_2} & \sim 0.040 \; M_{1/2}^2, \qquad
\widetilde{m}^2_{u_2} & \sim 0.072 \; M_{1/2}^2, \qquad
\widetilde{m}^2_{d_2} & \sim 0.025 \; M_{1/2}^2, \nonumber \\
\widetilde{m}^2_{q_3} & \sim 0.140 \; M_{1/2}^2, \qquad
\widetilde{m}^2_{u_3} & \sim 0.140 \; M_{1/2}^2, \qquad
\widetilde{m}^2_{d_3} & \sim 0.030 \; M_{1/2}^2,
\label{eq:squarkmasses}
\end{alignat}
which after applying the CKM rotations and evolving down to the
weak scale will result in, {\em i.e.},
\bea
\delta^{LL}_{sd} & \sim & \frac{\widetilde{m}^2_{s} - \widetilde{m}^2_{d}}
{ \widetilde{m}^2_{s} } \times V_{us} \sim 5 \times 10^{-4},
\eea
(with similar results for $\delta^{RR}$), acceptably small
\cite{Gabbiani:1996hi}.  And for the leptons we have,
\begin{alignat}{3}
\widetilde{m}^2_{l_1} & \sim 0.016 \; M_{1/2}^2, \qquad
\widetilde{m}^2_{e_1} & \sim 0.017 \; M_{1/2}^2, \qquad & \nonumber \\
\widetilde{m}^2_{l_2} & \sim 0.026 \; M_{1/2}^2, \qquad
\widetilde{m}^2_{e_2} & \sim 0.050 \; M_{1/2}^2, \qquad & \nonumber \\
\widetilde{m}^2_{l_3} & \sim 0.026 \; M_{1/2}^2, \qquad
\widetilde{m}^2_{e_3} & \sim 0.140 \; M_{1/2}^2 . \qquad &
\label{eq:sleptonmasses}
\end{alignat}
The much larger corrections to the masses of $q_3$, $u^c_3$, and $e^c_3$
are a direct result of those fields being localized around the Higgs brane,
and thus having wave functions concentrated closer to the supersymmetry breaking
brane.  The Higgses receive negligibly small soft masses at $M_c$.
Thus, we see that the flavor model has left an imprint of sorts
on the sparticle mass spectrum.

The resulting weak scale sparticle masses (with the $\mu$ term fixed by the
requirement of proper EWSB - for specific gaugino-mediation solutions to
the $\mu$-problem, see \cite{Kaplan:2000av,Chacko:2000mi,Schmaltz:2000gy})
show some distinct differences from standard gaugino-mediation.  First,
the lightest superpartner is typically a neutralino as opposed to
a stau, because of the additional contribution to stau masses in
Eq.~(\ref{eq:sleptonmasses}).  This feature allows us to more simply
connect with a standard picture of cosmology.  Further, there is
non-degeneracy of the squarks and sleptons of different families,
and for different chiralities,
with the most pronounced difference for the third generation.  This is
a direct consequence of the fact that the extra dimensions play a nontrivial
role both in generating flavor breaking and in supersymmetry breaking, and
represents a way in which future experiments could make progress
to unravel the flavor puzzle, by making precision measurements
of the supersymmetry-breaking parameters.

\section{Conclusion}
\label{conclusion}
\indent \indent
In this article we have examined a variety of tools, both in supersymmetric
and non-supersymmetric contexts, by which one can recover the spectrum
of fermion masses through the localization of matter fields in an extra dimension.
Orbifold boundary conditions allow us to complete the Arkani-Hamed-Schmaltz
model for the first time, resulting in a theory which actually
contains chiral matter.  Going further, we construct two new non-supersymmetric
models which successfully realize quark and lepton flavor from an
underlying theory containing only parameters of order one.
We have examined the constraints from flavor-changing neutral currents
and $CP$ violation as applied to the Kaon system, and find that our new
models relax the experimental bounds on the fundamental scale compared to those on
the AS model.

However, these constraints remain strong, requiring $M_* \gtrsim 10^3 M_W$,
and disfavor the use of large extra dimensions to explain both flavor and
the hierarchy problem.  Thus we consider supersymmetric theories, where
we can use much smaller extra dimensions, safe from flavor constraints
related to KK modes of the gauge bosons.  After reviewing the powerful
notation that expresses 5 dimensional supersymmetry in terms of superfields,
we find non-trivial flat directions which preserve $\N=1$ supersymmetry
and can themselves be used to localize chiral superfields.

These tools allow us to supersymmetrize our most successful
non-supersymmetric flavor model, by invoking an orbifold and odd mass terms
for hypermultiplets which result in chiral fields localized around the
orbifold fixed points.  One possible origin of this odd mass could be
from the $\N=2$ superpartners of the 5d gauge superfield for a
U(1) gauge group which is maximally broken.
Another interesting supersymmetric model is compactified without the orbifold,
and generates a chiral theory by projecting chiral brane fields into the bulk
through mixing with bulk hypermultiplets.  This allows us to consider
a supersymmetric version of the AS model, where the Higgses (as members
of hypermultiplets) as well as the fermions are localized across the
extra dimension.

If supersymmetry-breaking is also extra-dimensional, as in
gaugino-mediation, the fact that the extra dimension plays a dual
role can manifest itself in the low energy superpartner mass spectrum,
and allows one to see evidence for the mechanism of flavor by
carefully measuring the superparticle masses.  If instead supersymmetry
is broken by an exponentially small twist in the boundary conditions
({\it i.e.}, the Scherk-Schwarz mechanism),
our model successfully realizes a field theory mechanism for small
Yukawas without disrupting the flavor degeneracy of the sfermions.
Unfortunately, we do not know of any new weak-scale predictions in
this case.

To summarize, an extra dimension allows for a new perspective on many
of the puzzles in particle physics today.  In assembling specific
tools for one exciting feature - the localization of fields - it is our
hope that these will prove useful in building models that are on the one hand
beautiful and elegant, and on the other complete and realistic.

\section{Acknowledgements}
\indent \indent
The authors are grateful for conversations with N.~Arkani-Hamed,
H.--C.~Cheng, A.~Romanino, M.~Schmaltz, C.E.M~Wagner, and N.~Weiner.
Work at Argonne National Lab
is supported in part by the DOE under contract W-31-109-ENG-38.
Part of this work is also supported in part by the DOE under contracts
DE-FG02-90ER40560 and DE-AC03-76SF00515.


\end{document}